\documentclass[10pt]{article}
\usepackage{cite}
\usepackage{amsmath,amssymb,amsfonts}
\usepackage{algorithmic}
\usepackage{graphicx}
\usepackage{amsthm,bm,booktabs,hyperref,xcolor, cite}

\usepackage{textcomp}
\def\BibTeX{{\rm B\kern-.05em{\sc i\kern-.025em b}\kern-.08em
    T\kern-.1667em\lower.7ex\hbox{E}\kern-.125emX}}

\newcommand{\xx}{\hat{\mathbf x}}

\usepackage[margin=1in]{geometry}
    
\title{Spectral integral method for reflection and transmission by one-dimensional periodic arrays of penetrable cylinders}
\author{Ergun Simsek \\
\small{Department of Computer Science and Electrical Engineering} \\
\small{ University of Maryland Baltimore County} \\
\small{ 1000 Hilltop Circle, Baltimore, MD 21250, USA} \\
\small{E-mail: simsek@umbc.edu} }

\newcommand{\dn}{\partial_n}
\newcommand{\ve}{\varepsilon}
\newcommand{\rr}{\mathbf r}
\newcommand{\nn}{\hat{\mathbf n}}

\begin{document}
\maketitle

\begin{abstract}
The spectral integral method (SIM) is extended from a single cylinder to an infinite one-dimensional (1-D) periodic array (a grating) of identical, homogeneous, arbitrarily shaped smooth cylinders. The cylinders may be dielectric or metallic with complex electrical permittivity. Interactions among all members of the array are carried exactly by the 1-D quasi-periodic Green's function, $G_p$, in the exterior boundary integral equation, so only the boundary of one reference cylinder is discretized. $G_p$ is written as the free-space Green's function plus a regular part $\tilde G$ that is infinitely differentiable on the reference cylinder whenever the cylinders do not touch. $\tilde G$ and its gradient are evaluated with Ewald's method. Floquet reflection and transmission amplitudes, as well as reflectance, transmittance, and absorptance, are obtained from the boundary data. For the analytic boundaries and numerical examples considered here, the SIM exhibits root-exponential convergence in $N$ for glass and gold gratings, where $N$ is the number of samples on the boundary, for both polarizations, at oblique incidence and with several diffraction orders. Several numerical examples are provided.
\end{abstract}

\section{Introduction}\label{sec:intro}
Boundary integral methods for gratings have a long history. Maystre's integral theory \cite{maystre78} treats the grating profile as a periodic interface that may be a metal of finite conductivity, and N\'ed\'elec and Starling \cite{nedelec91} placed quasi-periodic integral equations on a rigorous footing. The kernel of these formulations is the quasi-periodic Green's function, a lattice sum of Hankel functions that converges too slowly to be summed directly, and its efficient evaluation has been studied through Ewald splitting \cite{jordan86, kustepeli, capolino}, lattice sums \cite{yasumoto99, linton, linton10}, and representations tailored to high frequencies \cite{kurkcu09}. These representations share one limitation: at a Rayleigh--Wood anomaly, one spectral term diverges and the Green's function is no longer mathematically well-defined. Near the anomaly, its evaluation loses accuracy. Multipole methods, which expand the field of each cylinder in cylindrical harmonics and couple the cylinders through lattice sums \cite{twersky, nicorovici95, botten00}, are very accurate for circular cylinders.

Two families of methods remain accurate at the anomalies. Barnett and Greengard \cite{barnett11} avoid the quasi-periodic Green's function altogether and enforce quasi-periodicity through auxiliary layer potentials on the walls of the unit cell. Their approach has been combined with fast direct solvers and extended to multilayered media and inclusions \cite{gillman13, cho15, lai15}. Bruno {\textit{et al.}} use smoothly windowed and shifted Green's functions that converge rapidly throughout the spectrum for periodic surfaces \cite{brunoHaslam09, brunoDelourme14} and for arrays of cylinders \cite{brunoFL17}. Both families rely on high-order singular quadratures and have been demonstrated mainly for impenetrable obstacles and dielectric scatterers. Plasmonic gratings have been modeled with surface integral equations \cite{gallinet10}, but with low-order boundary elements and therefore algebraic convergence. In a metal at optical frequencies, the interior wavenumber is complex, so the interior Green's function might vary by many orders of magnitude around the boundary of a cylinder, and the treatment of the logarithmic singularity must remain accurate over that whole range. This regime has been analyzed for single scatterers \cite{helsing18, simsekTE} but, to our knowledge, not for gratings.

The spectral integral method (SIM) offers a simpler route \cite{sim01}. It represents the boundary unknowns by trigonometric interpolation on equispaced nodes, evaluates the boundary convolutions with the fast Fourier transform, and removes the logarithmic singularity analytically in the Fourier domain, so that no specialized quadrature rule is needed. The original SIM \cite{sim01} and its extended versions \cite{simLM, sim3D} treated nonperiodic and periodic structures made of perfect electric conductors or lossless dielectrics, but they did not compute the individual Floquet amplitudes and did not address Rayleigh--Wood anomalies. A recent study \cite{simsekTE} showed that its global singularity treatment fails for large lossy scatterers because of catastrophic cancellation, and replaced it with a windowed singularity subtraction that restores exponential convergence.

Here, the windowed SIM is extended to gratings that are periodic in one direction and illuminated by a plane wave at an arbitrary angle of incidence, in both the TE$_z$ ($H_z$) and TM$_z$ ($E_z$) polarizations. The TE$_z$ polarization is the one relevant to plasmonics. Its electric field lies in the cross-sectional plane, has a component normal to the boundary, and induces surface charges, which support localized surface plasmons on metal cylinders and their diffractively coupled lattice modes in gratings \cite{simsekTE, maier, kravets, porto}. In this work, the periodicity enters through the Ewald representation of the quasi-periodic Green's function, evaluated through the Faddeeva function so that it remains valid for any angle of incidence and for complex arguments. The two polarizations are treated within a single formulation, in which they differ only through the ratio of the interior and exterior normal derivatives in the boundary conditions. The cross-section may be any smooth closed curve that fits within one period, and its height normal to the grating is unrestricted.

The formulation also admits complex permittivities with a negative real part, which is the regime of noble metals in the visible and near infrared, of doped semiconductors and transparent conducting oxides in the infrared. Such scatterers raise three difficulties. First, the interior field is confined to a thin skin layer, and the interior Green's function grows or decays exponentially across the object, which is what defeats the global singularity subtraction of \cite{sim01} and is resolved by the windowed one \cite{simsekTE}. Second, in TE$_z$ the jump of the normal derivative changes sign, and near the planar surface-plasmon condition the boundary integral system becomes nearly singular, so that the resonances of interest are governed by material loss rather than by geometry. Third, absorption becomes an output of the simulation alongside reflection and transmission, and energy conservation is no longer automatic. Hence, the Floquet amplitudes have to be computed for every propagating order, and the residual of the energy balance can be used as a diagnostic.

The method is verified in both polarizations against an independent multipole solution for gratings of circular cylinders \cite{twersky, linton}, and the two agree to nearly machine precision. Although the method is not defined exactly at a Rayleigh--Wood anomaly, its accuracy degrades only slowly as the anomaly is approached. We then show that a grating of gold wires acts as a polarizing beam splitter, and we reproduce the transmission resonances of the free-standing lamellar metallic grating of Porto \emph{et al.} \cite{porto}, whose near-rectangular bars are modeled as high-order superellipses.

The paper is organized as follows. The problem, the quasi-periodic Green's function, the boundary integral equations, their spectral discretization, and the extraction of the diffraction efficiencies are presented first. We then present numerical examples with cylinders made of lossless dielectrics and lossy metals, study the accuracy with respect to the sampling density, discuss the behavior near Rayleigh--Wood anomalies, and we conclude.

\section{Numerical Formulation}
\subsection{Spectral Integral Method for 1-D Periodic Structures}
\begin{figure}[htbp]
\begin{center}
\includegraphics[width=0.6\linewidth]{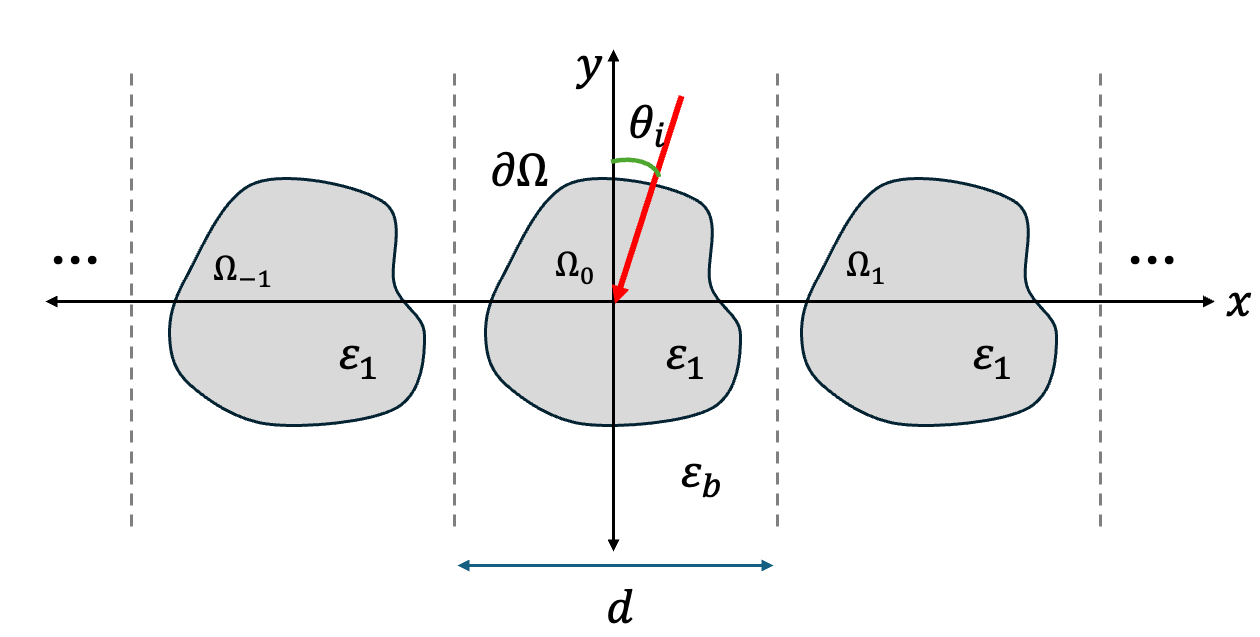}
\caption{Schematic illustration of one-dimensional periodic scatterers.}
\label{figure1}
\end{center}
\end{figure}

Identical cylinders, $\Omega_p=\Omega_0+p\,d\,\xx$, $p\in\mathbb Z$, with relative permittivity $\ve_1\in\mathbb C$ ($\operatorname{Im}\ve_1\ge0$) are embedded in a lossless background $\ve_b>0$, with $\mu=\mu_0$ everywhere, as shown in Fig.~\ref{figure1}. The cross-section $\Omega_0$ has a smooth boundary $\partial\Omega_0$ that lies inside the strip $|x|<d/2$, and $d$ is the grating periodic. The time dependence is $e^{-i\omega t}$; the wavenumbers of the background and of the cylinders are $k_b=k_0\sqrt{\ve_b}$ and $k_1=k_0\sqrt{\ve_1}$, respectively.

\noindent A plane wave of unit amplitude is incident from $y>0$,
\begin{equation}
U^{\rm inc}=e^{i(\beta_0 x-\gamma_0y)},\qquad \beta_0=k_b\sin\theta_i,\quad \gamma_0=k_b\cos\theta_i,
\label{eq:inc}
\end{equation}
so that its direction of propagation makes the angle $\theta_i$ with the $y$ axis. Here $U=H_z$ for TE$_z$ and $U=E_z$ for TM$_z$. The field is quasi-periodic, $U(x+d,y)=e^{i\beta_0 d}U(x,y)$, and for $|y|>y_{\max}$, it has the Rayleigh (Floquet) expansions
\begin{equation}
U=\begin{cases}
U^{\rm inc}+\displaystyle\sum_n R_n\,e^{i(\beta_n x+\gamma_n y)}, & y>y_{\max},\\[2ex]
\displaystyle\sum_n T_n\,e^{i(\beta_n x-\gamma_n y)}, & y<-y_{\max},
\end{cases}
\label{eq:rayleigh}
\end{equation}
where $y_{\max}$ is the maximum absolute vertical coordinate of the cylinder's boundary, $R_n$ and $T_n$ are the Floquet reflection and transmission coefficients for the mode number $n$, $\beta_n=\beta_0+{2\pi n}/{d}$, and $\gamma_n=\sqrt{k_b^2-\beta_n^2}$, with $\operatorname{Re}\{\gamma_n\}\ge0$ and $\operatorname{Im}\{\gamma_n\}\ge0$. Orders with $|\beta_n|<k_b$ propagate (diffraction orders), and the others are evanescent; $\gamma_n=0$ defines a Rayleigh--Wood anomaly \cite{wood, rayleigh, maystre, fem1}. The transmission conditions on $\partial\Omega_0$ are $U^+=U^-$ and $\dn U^-=\zeta\,\dn U^+$, where the superscripts $+$ and $-$ denote exterior and interior limits, and $\zeta=\ve_1/\ve_b$ for TE$_z$ and $\zeta=1$ for TM$_z$. The unit normal $\nn$ points out of $\Omega_0$, and $\partial\Omega_0$ is described by a counter-clockwise parametrization $\rr(t)$, $t\in[0,2\pi)$.

\noindent The quasi-periodic Green's function is the field of an array of line sources $\delta(\rr-\rr'-pd\xx)\,e^{i\beta_0 pd}$,
\begin{equation}
G_p(\rr,\rr')=\frac i4\sum_{p=-\infty}^{\infty}e^{i\beta_0 pd}\,H^{(1)}_0\!\big(k_b|\mathbf R_p|\big),
\label{eq:Gp}
\end{equation}
where $H^{(1)}_0(\cdot)$ is the zeroth-order Hankel function of the first kind, $\mathbf R=(X,Y)=\rr-\rr'$ and $ \mathbf R_p=\mathbf R-pd\,\xx$. The quasi-periodic Green's function satisfies
\begin{equation}
(\nabla^2+k_b^2)G_p=-\sum_p\delta(\mathbf R_p)\,e^{i\beta_0 pd},
\end{equation}
and it is quasi-periodic in $X$, and is outgoing as $|Y|\to\infty$. Poisson summation gives the spectral (Floquet) form
\begin{equation}
G_p=\frac{i}{2d}\sum_{n}\frac{e^{i\beta_nX+i\gamma_n|Y|}}{\gamma_n}
=\frac{1}{2d}\sum_n\frac{e^{i\beta_nX-\kappa_n|Y|}}{\kappa_n},
\label{eq:spec}
\end{equation}
where $\kappa_n=-i\gamma_n=\sqrt{\beta_n^2-k_b^2}$ with $\operatorname{Re}\kappa_n\ge0$. Neither form is suitable for evaluation on the boundary. For real $k_b$, the terms of Eq.~\eqref{eq:Gp} decay only like $|p|^{-1/2}$, so the series converges conditionally, and it diverges at Rayleigh--Wood anomalies. The terms of Eq.~\eqref{eq:spec} decay like $e^{-2\pi|n||Y|/d}$, so the series converges slowly as $Y\to0$, which is precisely the regime required by the boundary integral equations.

\noindent The Ewald representation is derived from the integral form of the free-space Green's function \cite{capolino,linton},
\begin{equation}
\frac{i}{4}H_0^{(1)}(kR)=\frac{1}{2\pi}\int_0^{\infty}\frac{1}{s} \exp\!\Big(-R^2s^2+\frac{k^2}{4s^2}\Big) ds,
\label{eq:ewaldint}
\end{equation}
where the path of integration is deformed into the complex $s$-plane so that the integrand decays at both endpoints. The integral is split at $s=\eta>0$, where $\eta$ is the Ewald splitting parameter \cite{kustepeli}. On the interval $s>\eta$, the factor $e^{k^2/4s^2}$ is expanded in a power series and the resulting series is integrated term by term. For real $k>0$, the integrand grows without bound as $s\to0$ along the real axis, so Eq.~\eqref{eq:ewaldint} is understood as a contour integral: the path leaves $s=0$ in the sector $-\pi/2<\arg s<-\pi/4$, where $\operatorname{Re}(k^2/s^2)<0$ and the integrand decays, and then returns to the positive real axis, along which it continues to $s=\infty$ \cite{linton,capolino}. This choice yields the outgoing function $H_0^{(1)}$ for the time dependence $e^{-i\omega t}$; with $k^2$ replaced by $-\kappa^2$, $\kappa>0$, the path may be taken along the real axis and Eq.~\eqref{eq:ewaldint} reduces to the familiar representation of $K_0(\kappa R)/2\pi$.  On the interval $s>\eta$, $|e^{k^2/4s^2}|\le e^{k^2/4\eta^2}$, so the factor $e^{k^2/4s^2}$ can be expanded in a power series and the resulting series integrated term by term. On the remaining part of the path, the sum over $p$ is first transformed by Poisson summation. Each Floquet order then contributes an integrand proportional to $s^{-2}\exp\!\big(-Y^2s^2-\kappa_n^2/4s^2\big)$, whose integral is available in closed form. For evanescent orders, $\kappa_n^2>0$ and the integrand decays as $s\to0$ along the real axis. For the finitely many propagating orders, $\kappa_n^2=-\gamma_n^2<0$; the contour is needed only for these, and its effect is accounted for exactly by evaluating the closed form at $\kappa_n=-i\gamma_n$ on the branch $\operatorname{Re}\kappa_n\ge0$. No integral is evaluated numerically near $s=0$. At a Rayleigh--Wood anomaly, $\kappa_n=0$ and the integrand $s^{-2}e^{-Y^2s^2}$ is not integrable at $s=0$, which is the origin of the $1/\kappa_n$ singularity discussed below.  On the interval $s<\eta$, the sum over $p$ is first transformed by Poisson summation \cite{benedetto97}. The result is $G_p=G_{\rm spec}+G_{\rm spat}$ with \cite{kustepeli,capolino}
\begin{align}
G_{\rm spec}&=\frac{1}{4d}\sum_n\frac{e^{i\beta_nX}}{\kappa_n}\Big[P(\kappa_n,Y)+P(\kappa_n,-Y)\Big],\label{eq:Gspec}\\
P(\kappa,\tau)&=e^{\kappa\tau}\operatorname{erfc}\!\Big(\frac{\kappa}{2\eta}+\tau\eta\Big),\label{eq:P} \\
G_{\rm spat}&=\frac1{4\pi}\sum_pe^{i\beta_0 pd}\sum_{q=0}^\infty\alpha_q\,\Upsilon_{q+1}\big(|\mathbf R_p|^2\eta^2\big), \label{eq:Gspat} \\
\alpha_q & =\frac{1}{q!}\Big(\frac{k_b}{2\eta}\Big)^{2q},
\end{align}
where $\tau$ is a real variable, which takes the values $\pm Y$ in Eq.~\eqref{eq:Gspec}, and $\Upsilon_{q+1}(z)=\int_1^\infty e^{-zt}t^{-q-1}\,dt$ is the generalized exponential integral. Both series converge like Gaussians, since $\operatorname{erfc}(\kappa_n/2\eta)\sim e^{-(\pi n/\eta d)^2}$ and $\Upsilon_1(|\mathbf R_p|^2\eta^2)\sim e^{-(pd\eta)^2}$. In the limit $\eta\to\infty$, $G_{\rm spat}\to0$ and Eq.~\eqref{eq:Gspec} reduces to Eq.~\eqref{eq:spec}. The Ewald representation accelerates the convergence of the lattice sum \cite{capolino}, but it does not remove the singularity of $G_p$ at a Rayleigh--Wood anomaly. As $\kappa_n\to0$, $P(\kappa_n,Y)+P(\kappa_n,-Y)\to\operatorname{erfc}(\eta Y)+\operatorname{erfc}(-\eta Y)=2$, so the $n$-th term of Eq.~\eqref{eq:Gspec} grows like $1/\kappa_n$. The formulation therefore requires $\gamma_n\neq0$ for all $n$; Section~\ref{sec:wood} shows that it remains accurate down to $|\kappa_n|d\approx3\times10^{-4}$.

\noindent The gradients follow from $d\Upsilon_{q+1}(z)/dz=-\Upsilon_q(z)$, with $\Upsilon_0(z)=e^{-z}/z$, and from the cancellation of the Gaussian terms in $\partial_Y\big[P(\kappa,Y)+P(\kappa,-Y)\big]$, where
\begin{align}
\partial_\tau P(\kappa,\tau)&=\kappa P(\kappa,\tau)-(2\eta/\sqrt\pi)\,e^{-\kappa^2/4\eta^2-\tau^2\eta^2} \\
\partial_XG_{\rm spec}&=\frac1{4d}\sum_n\frac{i\beta_n\,e^{i\beta_nX}}{\kappa_n}\Big[P(\kappa_n,Y)+P(\kappa_n,-Y)\Big],\\
\partial_YG_{\rm spec}&=\frac1{4d}\sum_ne^{i\beta_nX}\Big[P(\kappa_n,Y)-P(\kappa_n,-Y)\Big],\\
\nabla G_{\rm spat}&=-\frac{\eta^2}{2\pi}\sum_pe^{i\beta_0 pd}\,\mathbf R_p\sum_{q=0}^\infty\alpha_q\,\Upsilon_q\big(|\mathbf R_p|^2\eta^2\big).
\end{align}

\noindent For propagating orders, $\kappa_n=-i\gamma_n$ is imaginary, so the error function must be evaluated for complex arguments, for which real-argument implementations cannot be used. With $u=\kappa/2\eta+\tau\eta$,
\begin{equation}
P(\kappa,\tau)=
\begin{cases}
e^{-\kappa^2/4\eta^2-\tau^2\eta^2}\,w(iu), & \operatorname{Re}u\ge0,\\[1ex]
2e^{\kappa\tau}-e^{-\kappa^2/4\eta^2-\tau^2\eta^2}\,w(-iu), & \operatorname{Re}u<0,
\end{cases}
\label{eq:Pfaddeeva}
\end{equation}
where $w(z)=e^{-z^2}\operatorname{erfc}(-iz)$ is the Faddeeva function \cite{Faddeeva, weideman}. It is evaluated with Weideman's rational approximation \cite{weideman}, using 40 terms. Compared with an independent double-precision implementation, its relative error is below $1.6\times10^{-14}$ throughout the closed upper half-plane for $|z|\le60$. In both cases of Eq.~\eqref{eq:Pfaddeeva}, the argument of $w$ lies in the closed upper half-plane, where $|w|\le1$, and the exponential prefactors remain bounded. For propagating orders, $e^{-\kappa_n^2/4\eta^2}=e^{\gamma_n^2/4\eta^2}\le e^{k_b^2/4\eta^2}\le e^{\nu^2}$ with the choice of $\eta$ discussed at the end of Section \ref{sec:convN}, and for evanescent orders this factor does not exceed one. The term $2e^{\kappa\tau}$ is used only when $\operatorname{Re}u<0$, which implies $|e^{\kappa\tau}|\le1$. Equation~\eqref{eq:Pfaddeeva} therefore does not overflow. 

\noindent For $\rr,\rr'\in\partial\Omega_0$ we have $|X|<d$, so only the $p=0$ term of Eq.~\eqref{eq:Gp} is singular. We therefore define the regular part
\begin{equation}
\tilde G(\mathbf R)=G_p-\frac i4H_0^{(1)}(k_bR)
=G_{\rm spec}+\frac1{4\pi}\sum_{p\neq0}e^{i\beta_0 pd}\sum_{q=0}^\infty\alpha_q\,\Upsilon_{q+1}\big(|\mathbf R_p|^2\eta^2\big)+B(R),
\label{eq:Gtilde}
\end{equation}
with
\begin{equation}
B(R)=\frac1{4\pi}\sum_{q=0}^\infty\alpha_q\,\Upsilon_{q+1}(R^2\eta^2)-\frac i4H^{(1)}_0(k_bR).
\end{equation}
Because $\tilde G$ equals the sum over the images $p\neq0$, it is analytic for $|X|<d$. Using $\Upsilon_1(z)=-\gamma_{\rm E}-\ln z+O(z)$, where $\gamma_{\rm E}$ is the Euler--Mascheroni constant \cite{EMconstant}, $\Upsilon_{q+1}(0)=1/q$ for $q\ge1$, and the small-argument form of $H_0^{(1)}$, the logarithmic singularities cancel and
\begin{equation}
\lim_{R\to0}B(R)=\frac{1}{2\pi}\ln\frac{k_b}{2\eta}+\frac{\gamma_{\rm E}}{4\pi}+\frac{1}{4\pi}\sum_{q\ge1}\frac{\alpha_q}{q}-\frac i4 .
\end{equation}
$B$ is smooth and radially symmetric, so $\nabla B=0$ at $R=0$. For $R>0$,
\begin{equation}
\nabla B=-\frac{\eta^2}{2\pi}\,\mathbf R\sum_{q=0}^\infty\alpha_q\,\Upsilon_q(R^2\eta^2)+\frac{ik_b}{4}H_1^{(1)}(k_bR)\,\frac{\mathbf R}{R},
\end{equation}
in which the $1/R$ singularities of the two terms cancel analytically and $H_1^{(1)}(\cdot)$ is the first-order Hankel function of the first kind. Numerically, $\tilde G$ is continuous to 10 digits down to $R=10^{-6}d$. Its gradient, which enters the double-layer kernel, is equally well behaved: the deviation of $\nabla\tilde G$ from its value at $R=0$ decreases linearly with $R$ down to $R=10^{-7}d$, and at $R=0.02d$ the computed gradient agrees with a central finite-difference derivative of $\tilde G$ to $2\times10^{-8}$, the accuracy of the difference formula.

Then, we apply Green's second identity to the scattered field $U^s=U-U^{\rm inc}$ and $G_p$ in the part of the unit cell $|x|<d/2$ outside $\Omega_0$, truncated at $|y|=h$, with $h\to\infty$. The contributions of the two side walls cancel: $U^s$ acquires the factor $e^{i\beta_0 d}$ and $G_p(\rr,\rr')$ the factor $e^{-i\beta_0 d}$ when $\rr'$ is shifted by $d\xx$, so the integrand is periodic, while the outward normals of the two walls are opposite. The contributions of the top and bottom boundaries vanish because both functions are superpositions of outgoing Floquet modes there. Since $U^{\rm inc}$ is regular inside $\Omega_0$, Green's second identity applied to $U^{\rm inc}$ and $G_p$ in $\Omega_0$ gives $\oint_{\partial\Omega_0}\big[U^{\rm inc}\partial_{n'}G_p-G_p\,\partial_{n'}U^{\rm inc}\big]ds'=0$ for $\rr$ outside $\Omega_0$. Adding this result and letting $\rr$ approach $\partial\Omega_0$ from outside gives the exterior equation
\begin{equation}
\tfrac12U(\rr)=U^{\rm inc}(\rr)+\oint_{\partial\Omega_0}\Big[U\,\frac{\partial G_p}{\partial n'}-G_p\,\dn U^+\Big]ds',\qquad \rr\in\partial\Omega_0 ,
\label{eq:extp}
\end{equation}
where the integral is a Cauchy principal value. The interior field involves only the material of $\Omega_0$, so the interior equation is the same as for a single cylinder, with $G_1=\frac i4H_0^{(1)}(k_1R)$:
\begin{equation}
\tfrac12U(\rr)=\oint_{\partial\Omega_0}\Big[\zeta\,G_1\,\dn U^+-U\,\frac{\partial G_1}{\partial n'}\Big]ds' ,\qquad \rr\in\partial\Omega_0 .
\label{eq:intp}
\end{equation}
All interactions between the cylinders of the array are therefore carried exactly by $G_p$, and only $\partial\Omega_0$ is discretized.

The boundary is sampled at the nodes $t_j=2\pi j/N$, $j=0,\dots,N-1$, and the unknowns are the vectors $\bm\Phi$ and $\bm\Psi$ of the nodal values of $U$ and $\dn U^+$. The discretization follows the single-cylinder SIM \cite{sim01}. Each kernel is split into a logarithmically singular part, whose convolution with the trigonometric interpolant of the density is evaluated exactly with the FFT, and a smooth remainder, which is integrated with the trapezoidal rule. With $G_p=G_b+\tilde G$, where $G_b=\frac i4H_0^{(1)}(k_bR)$ is the free-space Green's function of the background, and $\partial_{n'}\tilde G=-\nn'\cdot\nabla_{\mathbf R}\tilde G$, the smooth remainders of the exterior kernels become
\begin{align}
\bar G_b^{\,p}(t_j,t')&=\bar G_b(t_j,t')+\tilde G\big(\rr(t_j)-\rr(t')\big),\\
\bar H_b^{\,p}(t_j,t')&=\bar H_b(t_j,t')+\partial_{n'}\tilde G\big(\rr(t_j)-\rr(t')\big),
\end{align}
where $\bar G_b$ and $\bar H_b$ are the smooth remainders of the free-space kernels \cite{simsekTE}. Since $\tilde G$ is analytic on $\partial\Omega_0\times\partial\Omega_0$, the periodic remainders are also smooth, and the treatment of the logarithmic singularities is unchanged. The resulting linear system has the same structure as for a single cylinder,
\begin{align}
\mathbf C^b_p\bm\Phi+\mathbf D^b_p\bm\Psi & =\mathbf F,\\
\mathbf C^1\bm\Phi+\zeta\,\mathbf D^1\bm\Psi & =\mathbf 0,
\end{align}
where $F_j=U^{\rm inc}\big(\rr(t_j)\big)$, $\mathbf C^b_p$ and $\mathbf D^b_p$ are the exterior matrices with the periodic remainders, and $\mathbf C^1$ and $\mathbf D^1$ are the interior matrices. 

Finally, we calculate reflection, transmission, and absorption by 1-D periodic structures as follows. Below, the amplitudes $R_n$ and $T_n$ in Eq.~\eqref{eq:rayleigh} are the complex amplitudes of the $n^{\rm th}$ reflected and transmitted diffraction orders for an incident plane wave of unit amplitude. They follow directly from the boundary solution. Outside the cylinders, the scattered field $U^s=U-U^{\rm inc}$ is given by Green's representation formula,
\begin{equation}
U^s(\rr)=\oint_{\partial\Omega_0}\big[U\,\partial_{n'}G_p(\rr,\rr')-G_p(\rr,\rr')\,\psi\big]\,ds',
\label{eq:greenrep}
\end{equation}
where $U$ and $\psi=\dn U^+$ are the exterior boundary values of the total field and its normal derivative at $\rr'\in\partial\Omega_0$. For $y>y_{\max}$, all sources satisfy $y'<y$, and Eq.~\eqref{eq:spec} gives
\begin{equation}
G_p=\sum_n\frac{i}{2d\gamma_n}\,e^{i(\beta_nx+\gamma_ny)}\,e^{-i(\beta_nx'+\gamma_ny')},
\label{gp1}
\end{equation}
so that Eq.~\eqref{eq:greenrep} becomes a superposition of the upgoing Floquet modes $e^{i(\beta_nx+\gamma_ny)}$. Using $\partial_{n'}e^{-i(\beta_nx'+\gamma_ny')}=-i(\beta_nn'_x+\gamma_nn'_y)\,e^{-i(\beta_nx'+\gamma_ny')}$ and equating the coefficient of each mode with the corresponding term of Eq.~\eqref{eq:rayleigh} yields
\begin{equation}
R_n=\frac{i}{2d\gamma_n}\oint_{\partial\Omega_0}\Big[-i(\beta_nn'_x+\gamma_nn'_y)\,U-\psi\Big]e^{-i(\beta_nx'+\gamma_ny')}\,ds'.
\label{eq:Rn}
\end{equation}
For $y<-y_{\max}$, the same procedure with the downgoing modes $e^{i(\beta_nx-\gamma_ny)}$ gives the scattered part of the transmitted amplitudes. Adding the incident wave, which contributes only to the zeroth order, gives
\begin{equation}
T_n=\delta_{n0}+\frac{i}{2d\gamma_n}\oint_{\partial\Omega_0}\Big[-i(\beta_nn'_x-\gamma_nn'_y)\,U-\psi\Big]e^{-i(\beta_nx'-\gamma_ny')}\,ds'.
\label{eq:Tn}
\end{equation}

These integrals are evaluated with the trapezoidal rule in $t$, which is spectrally accurate for smooth periodic integrands. The power carried by each diffraction order follows from the period-averaged Poynting flux through a line $y=\mathrm{const}$. Normalizing every flux by the incident power crossing one period of this line, which is proportional to $\gamma_0 d$, gives the reflectance, transmittance and absorptance,
\begin{equation}
\mathcal R=\sum_{n\,\rm prop}\frac{\gamma_n}{\gamma_0}|R_n|^2,\qquad
\mathcal T=\sum_{n\,\rm prop}\frac{\gamma_n}{\gamma_0}|T_n|^2,\qquad
\mathcal A=\frac{C_{\rm abs}}{d\cos\theta_i},
\label{eq:RTA}
\end{equation}
where the sums run over the propagating orders, for which $\gamma_n$ is real. The absorption cross-section of one cylinder of the array,
\begin{equation}
C_{\rm abs}=-\frac{1}{k_b}\oint_{\partial\Omega_0}\operatorname{Im}(U^*\psi)\,ds,
\label{eq:Cabs}
\end{equation}
where the asterisk denotes complex conjugation, is the net power flowing into the cylinder per unit length along $z$, divided by the incident intensity, and has the dimension of length \cite{simsekTE}. Since the incident power crossing one period equals the incident intensity times $d\cos\theta_i$, $\mathcal A$ is the fraction of the incident power absorbed per period, and $\mathcal R$, $\mathcal T$ and $\mathcal A$ share the same normalization. Energy conservation then requires $\mathcal R+\mathcal T+\mathcal A=1$. Because $\mathcal A$ is computed from the boundary fields and $\mathcal R$ and $\mathcal T$ from the diffracted orders, the residual $1-\mathcal R-\mathcal T-\mathcal A$ provides a reference-free consistency check, which vanishes for the exact solution. This is--of course--nice to check but not a sufficient condition for accuracy, because errors in different quantities can compensate. 

\section{Numerical Results}\label{sec:acc}
We consider two types of materials: BK7 glass is a lossless dielectric, and gold (Au) is a lossy conductor with a negative electrical permittivity in the visible part of the electromagnetic spectrum. The wavelength-based electrical permittivity models and the parameters used are given in the supplementary document.

\subsection{Comparison with the Multipole Method}
We first compare the SIM with the multipole method (MPM) \cite{twersky,linton} for gratings of circular cylinders.  The MPM for gratings of circular cylinders is described in the supplementary document. In this method, the field scattered by each cylinder is expanded in outgoing cylindrical harmonics $H_l^{(1)}(k_b\rho)\,e^{il\phi}$ about its center, and quasi-periodicity relates the expansion coefficients of all cylinders to those of a single cylinder. The field incident on that cylinder is the plane wave plus the fields scattered by all other cylinders of the array, which are re-expanded about its center with Graf's addition theorem (Sec. 7.6.2 of \cite{erdelyi53}) through the lattice sums of the array. Imposing the transmission conditions through the analytical Mie coefficients of a circular cylinder then gives a linear system for the multipole coefficients $b_l$, $-L\le l\le L$, where $L$ is the truncation order of the expansion. The diffraction amplitudes follow in closed form from these coefficients. The MPM thus requires no boundary discretization and shares only the lattice sums with the SIM. Its accuracy is controlled by $L$: for all cases considered here, $L=60$ converges the amplitudes to the rounding level.

Two configurations are considered: a dielectric (N-BK7) grating and a gold grating studied as a function of wavelength and incidence angle. In both cases, the background is vacuum ($\ve_b=1$), the SIM uses $N=128$ boundary points, and both polarizations are computed.

\begin{figure}[htbp]
\centering
\includegraphics[width=\linewidth]{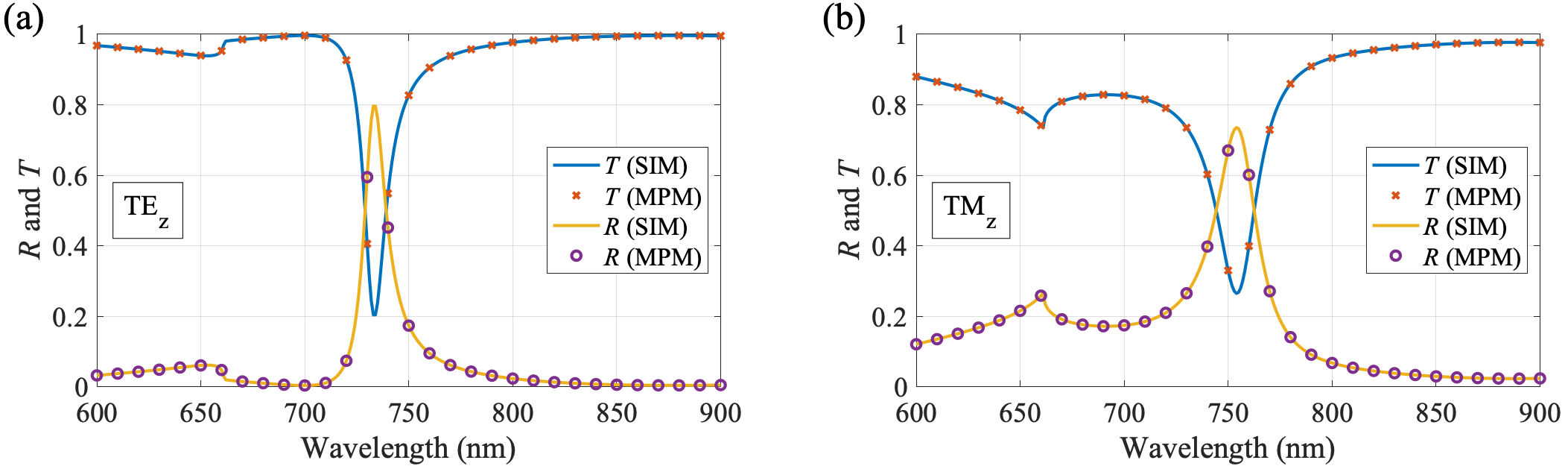}
\caption{Reflectance $\mathcal R$ and transmittance $\mathcal T$ of an N-BK7 cylinder grating ($a=200$ nm, $d=800$ nm) in vacuum at $\theta_i=10^\circ$, computed with the SIM ($N=128$, lines) and the MPM (symbols, every tenth wavelength shown). (a) TE$_z$. (b) TM$_z$.}
\label{fig:glassRT}
\end{figure}

\paragraph{N-BK7 grating.} Cylinders of radius $a=200$ nm are arranged with period $d=800$ nm ($2a/d=0.5$) and illuminated at $\theta_i=10^\circ$ for wavelengths from 600 to 900 nm in steps of 1 nm. The permittivity varies from $2.299$ at 600 nm to $2.277$ at 900 nm. The spectra for the TE$_z$ and TM$_z$ cases are shown in Fig.~\ref{fig:glassRT} (a) and (b), respectively. The kink at $661$ nm is the Rayleigh--Wood anomaly of the $n=+1$ order, $\lambda=d(1-\sin\theta_i)=661.1$ nm; above this wavelength only the orders $n=0$ and $n=-1$ propagate. A narrow resonance appears at 733 nm for TE$_z$ and at 754 nm for TM$_z$, where the reflectance reaches $0.796$ and $0.735$, respectively. Since the glass is lossless, $\mathcal R+\mathcal T=1$, and the SIM energy residual is below $1\times10^{-14}$ at all wavelengths. Over the full spectrum, the SIM and MPM values of $\mathcal R$ and $\mathcal T$ differ by at most $5.0\times10^{-14}$ (TE$_z$) and $3.8\times10^{-14}$ (TM$_z$), with the largest differences occurring at the resonances.

\begin{figure}[htbp]
\centering
\includegraphics[width=\linewidth]{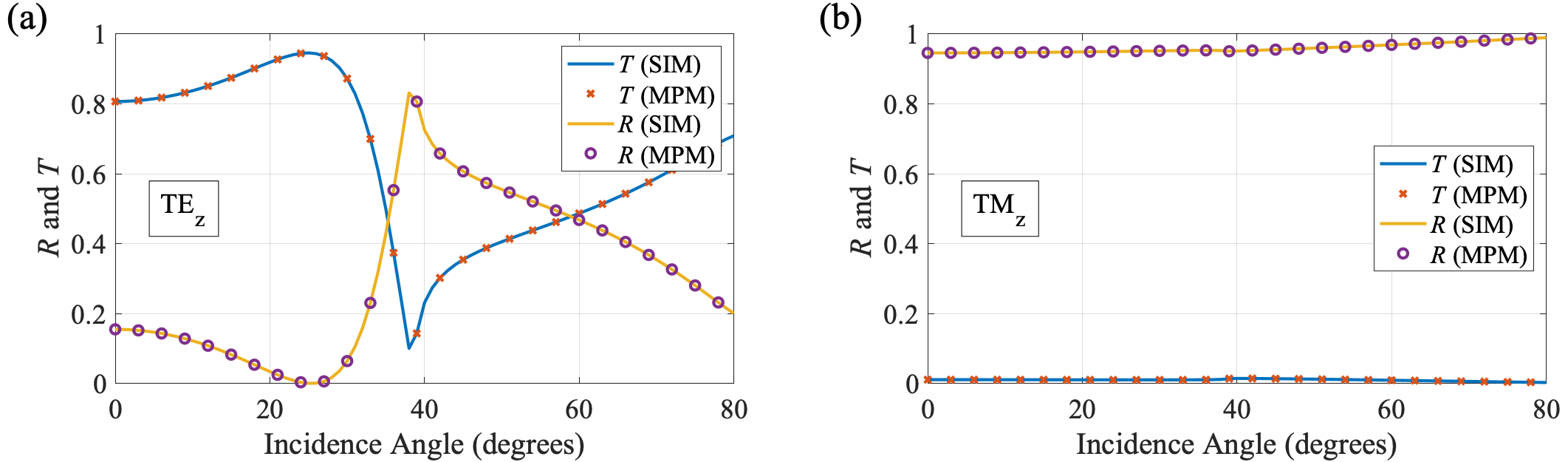}
\caption{Reflectance $\mathcal R$ and transmittance $\mathcal T$ of a gold cylinder grating ($a=130$ nm, $d=400$ nm) in vacuum at $\lambda=650$ nm as a function of the angle of incidence, computed with the SIM ($N=128$, lines) and the MPM (symbols, every third angle shown). (a) TE$_z$. (b) TM$_z$.}
\label{fig:goldRT}
\end{figure}

\paragraph{Gold grating.} Gold cylinders of radius $a=130$ nm are arranged with period $d=400$ nm ($2a/d=0.65$) and illuminated at $\lambda=650$ nm, where the Etchegoin model \cite{etchegoin} gives $\ve_1=-12.46+0.777i$. The angle of incidence is varied from $0^\circ$ to $80^\circ$ in steps of $1^\circ$. The results are shown in Fig.~\ref{fig:goldRT}. In TE$_z$, the reflectance first decreases to $4\times10^{-4}$ at $25^\circ$ and then rises sharply to $0.831$ at $38^\circ$, just below the Rayleigh--Wood anomaly of the $n=-1$ order, $\sin\theta_i=\lambda/d-1$, i.e.\ $\theta_i=38.68^\circ$, where the transmittance falls to $0.1$. In TM$_z$, the grating behaves as a wire-grid reflector, with $\mathcal R$ increasing from $0.946$ at normal incidence to $0.989$ at $80^\circ$ and $\mathcal T<0.015$ at all angles. The SIM and MPM values of $\mathcal R$ and $\mathcal T$ differ by at most $1.1\times10^{-13}$ (TE$_z$) and $5.4\times10^{-14}$ (TM$_z$) over all angles, and the SIM energy residual is below $3\times10^{-14}$.
Notice that this gold grating acts as a polarization-sensitive filter. At $\lambda=650$ nm and $\theta_i=25^\circ$, TM$_z$ light is almost completely reflected, with $\mathcal R\approx0.95$, while 95\% of the TE$_z$ light is transmitted, corresponding to a 33 dB reflection contrast.

In both configurations, the curves of the two methods are indistinguishable, and the differences between them are close to the level of double-precision rounding, even though the SIM uses only 128 boundary points and the two methods share only the lattice sums. The agreement holds for a dielectric and for a plasmonic material, for both polarizations, and across Rayleigh--Wood anomalies and resonances.

\subsection{Convergence with the number of boundary points}\label{sec:convN}
The convergence of the SIM is studied for the five gratings of circular cylinders (either BK7 or Au) listed in Table~\ref{tab:cases}, for which the MPM provides an independent reference. Filling fractions range from $2a/d=0.25$ to $0.91$ and incidence angles from $0^\circ$ to $30^\circ$. Case C2 has five propagating orders, C4 has three, and all other cases have only the zeroth order. The background is vacuum. For each case and polarization, the SIM is run with $N=16$ to $256$ boundary points, and the error is defined as the largest deviation of the complex diffraction amplitudes from the MPM reference over all propagating orders, $\max_n\big(|R_n-R_n^{\rm MPM}|,|T_n-T_n^{\rm MPM}|\big)$. The MPM reference uses $L=60$.

\begin{table}[h]\centering\footnotesize
\caption{Test gratings ($\lambda = 633$ nm, circular cylinders of radius $a$, period $d$).}\label{tab:cases}
\begin{tabular}{llrrcrrl}\toprule
Case & Material & $a (nm)$ & $d (nm)$ & $2a/d$ & $\theta_i$ ($^\circ$) \\\midrule
C1 & Glass & 200 & 500 & 0.80 & 15 \\
C2 & Glass & 300 & 1500 & 0.40 & 20  \\
C3 & Gold & 100 & 500 & 0.40 & 15  \\
C4 & Gold & 150 & 1200 & 0.25 & 30  \\
C5 & Gold & 100 & 220 & 0.91 & 0  \\
\bottomrule\end{tabular}\end{table}

\begin{figure}[htbp]
\centering
\includegraphics[width=0.9\linewidth]{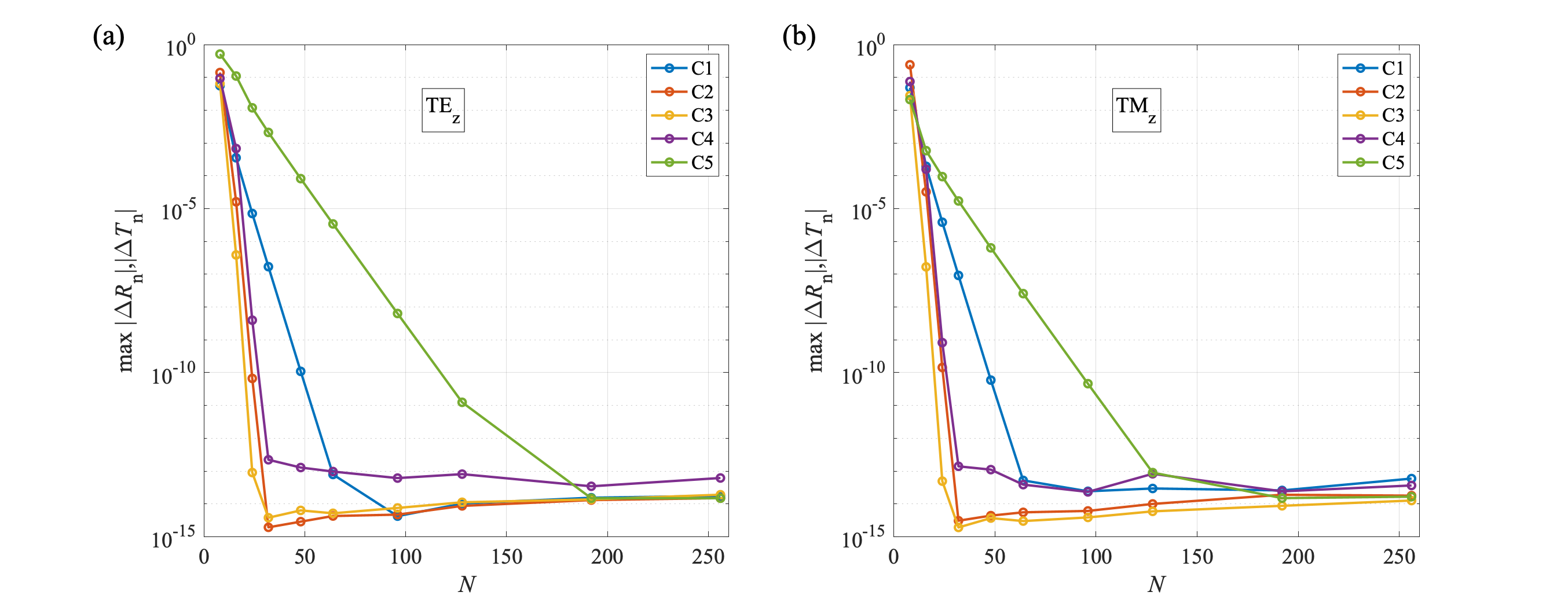}
\caption{Maximum error of the SIM diffraction amplitudes, $\max_n(|\Delta R_n|,|\Delta T_n|)$, relative to the MPM reference ($L=60$) versus the number of boundary points $N$ for the gratings of Table~\ref{tab:cases}. (a) TE$_z$. (b) TM$_z$.}
\label{fig:accuracy}
\end{figure}

\noindent The results are shown in Fig.~\ref{fig:accuracy}. In all cases, and in both polarizations, the error decreases root-exponentially with $N$ until it reaches a floor of $10^{-13}$--$10^{-14}$, set by double-precision rounding. Beyond this point, the error grows only slowly with $N$, because rounding errors accumulate in the larger linear system. For well-separated cylinders, the convergence is very fast. Cases C2 and C3 reach $2\times10^{-15}$ and $1\times10^{-14}$, respectively, with only $N=32$, and case C4, with three propagating orders, reaches $1\times10^{-13}$ at the same $N$. Case C1, whose glass cylinders fill 80\% of the period, requires $N=64$ for an error of $7\times10^{-14}$. These results hold equally for the lossless dielectric and for gold, whose negative permittivity does not slow down the convergence.

The rate of convergence is governed mainly by two geometric factors. The first is the gap $g=d-2a$ between neighboring cylinders. Because the field varies rapidly in a narrow gap, the number of points needed grows with the ratio of the perimeter to the gap. For C5 (gold, $g=20$ nm), the TE$_z$ error is $1.2\times10^{-11}$ at $N=128$ and $2\times10^{-14}$ at $N=256$. 

Here, we take $\eta=\max(\sqrt\pi/d,\ k_b/2\nu)$, with $\nu=1.5$ in all numerical results. The choice $\sqrt\pi/d$ is optimal at low frequencies \cite{kustepeli}. At high frequencies, the bound $k_b/2\eta\le\nu$ limits the magnitude of each part of the Ewald sum to about $e^{\nu^2}|G_p|$, so cancellation between the spectral and spatial parts costs at most $\nu^2/\ln10\approx1$ significant digit. Larger values of $\nu$ reduce the number of spectral terms at the price of more cancellation, and $\nu=1.5$ balances the two. This estimate is heuristic, but the results are insensitive to the choice: for case C4, the SIM--MPM discrepancy is less than  $10^{-12}$ for $\nu=0.75$, $1.5$ and $3$. The exponential integrals are computed with the upward recurrence $\Upsilon_{q+1}(z)=\big(e^{-z}-z\,\Upsilon_q(z)\big)/q$, starting from $\Upsilon_1(z)$, which is evaluated directly, and $\Upsilon_0(z)=e^{-z}/z$. Cancellation in this recurrence occurs only for large $z$, where all terms are of order $e^{-z}$ and therefore negligible. For $k_b/2\eta=\nu$, $10^{-10}\le z\le45$ and $q\le40$, the sums $\sum_q\alpha_q\Upsilon_{q+1}(z)$ and $\sum_q\alpha_q\Upsilon_q(z)$ agree with an independent implementation of the exponential integrals to within a relative error of $\sim\,10^{-15}$. For $|Y|\ge0.15d$, the Ewald values of $G_p$ and $\partial_YG_p$ agree with the direct spectral sum, Eq. \eqref{eq:spec}, (801 terms) to $8\times10^{-16}$. For an artificially lossy wavenumber $k=k_b(1+0.05i)$, for which Eq.~\eqref{eq:Gp} converges absolutely (6001 terms), $G_p$ and $\partial_XG_p$ agree with the direct spatial sum to $2\times10^{-15}$.

\subsection{Effect of the gap between neighbouring cylinders}\label{sec:gap}
\begin{figure}[htbp]
\centering
\includegraphics[width=0.8\linewidth]{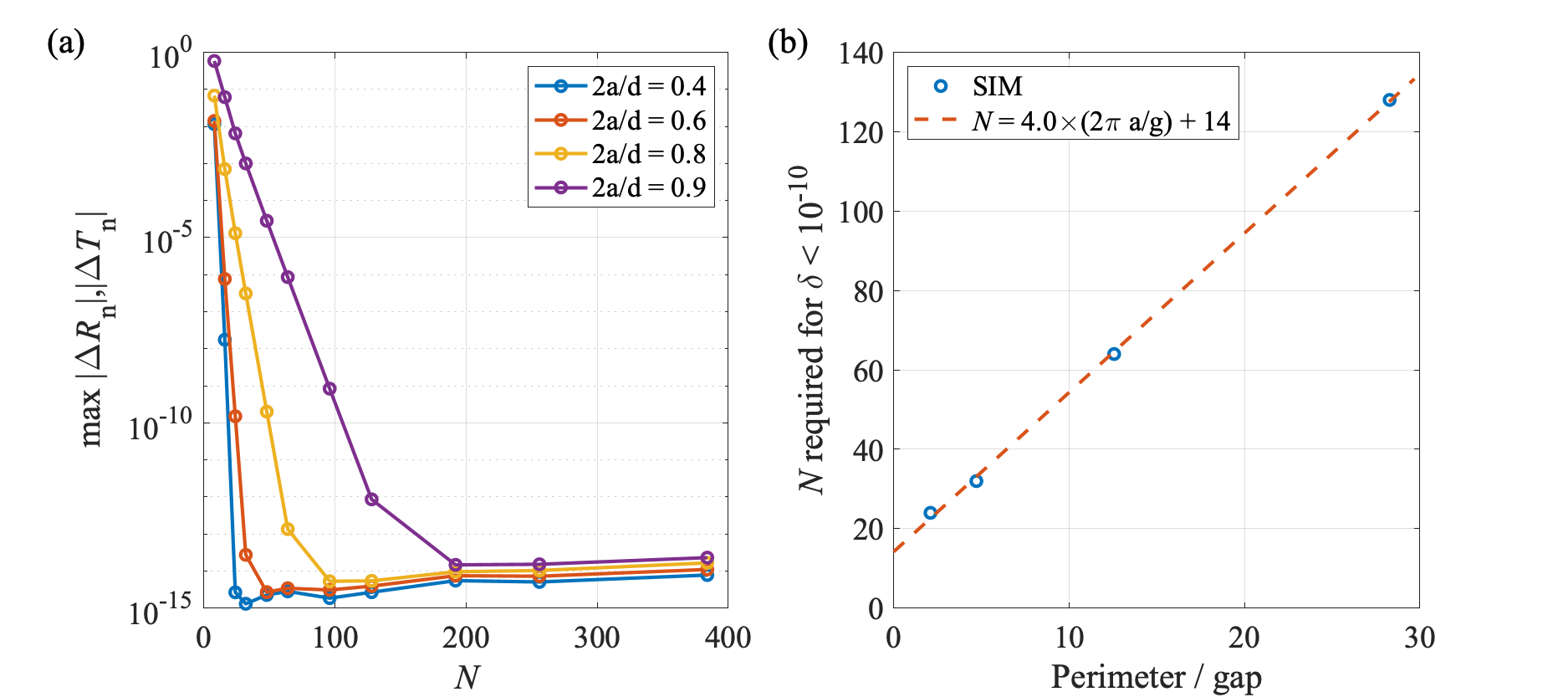}
\caption{(a) Maximum relative vs. $N$  for gold cylinders ($a=100$ nm, $\lambda=633$ nm, TE$_z$) with filling fractions $2a/d=0.4$, $0.6$, $0.8$ and $0.9$. (b) Number of boundary points $N$ required for an error $\delta<10^{-10}$ versus the ratio of the perimeter to the gap, $2\pi a/g$. The dashed line is the least-squares fit $N=4\times\,(2\pi a/g)+14$.}
\label{fig:gap}
\end{figure}

The convergence studies of the previous section indicate that narrow gaps between neighboring cylinders slow down the convergence of the SIM. To quantify this effect, we consider gold cylinders of radius $a=100$ nm at $\lambda=633$ nm in TE$_z$ polarization, and vary the period so that the filling fraction is $2a/d=0.4$, $0.6$, $0.8$ and $0.9$. The corresponding gaps are $300$, $133$, $50$ and $22$ nm, and the ratio of the perimeter to the gap, $2\pi a/g$, takes the values $2.09$, $4.71$, $12.57$ and $28.27$. Figure \ref{fig:gap} (a) shows the maximum error as a function of $N$. For each case, we determine the smallest number of boundary points in the test set $N\in\{16,24,32,48,64,96,128,192,256,384\}$ for which the error relative to the MPM reference, $\delta=\max_n\big(|\Delta R_n|,|\Delta T_n|\big)$, falls below $10^{-10}$.

The required numbers of points are $N=24$, $32$, $64$ and $128$, respectively (Fig.~\ref{fig:gap} (b)). They grow linearly with the perimeter-to-gap ratio, and a least-squares fit gives
\begin{equation}
N\approx4\times\,\frac{2\pi a}{g}+14 ,
\label{eq:Ngap}
\end{equation}
which reproduces all four values to within two points. The two terms have a simple interpretation. The constant term, $N\approx14$, is the number of points needed to resolve the field on an isolated cylinder of this size. The linear term states that the arc length between neighboring boundary points, $2\pi a/N$, does not have to exceed about one quarter of the gap. In the narrow gaps, the field of each cylinder varies on the length scale $g$ along the part of the boundary facing its neighbor, and this variation must be resolved by the equispaced nodes. Since the nodes are distributed uniformly in the parameter $t$, the resolution required in the gap determines $N$ for the whole boundary. Equation~\eqref{eq:Ngap} provides a practical rule for choosing $N$ in closely spaced gratings. Because the required $N$ is determined on a discrete set of values, the numbers in Fig.~\ref{fig:gap} are upper bounds, accurate to within the spacing of the test set.

\subsection{Accuracy at Rayleigh--Wood anomalies}\label{sec:wood}
At a Rayleigh--Wood anomaly, a diffraction order becomes grazing, $\gamma_n\to0$ \cite{wood,rayleigh,maystre}. The quasi-periodic Green's function, Eq. \eqref{eq:spec}, then contains a term $\propto1/\kappa_n$ that diverges, and the reflection and transmission coefficients have a square-root branch point: they are continuous, but their derivatives with respect to $\lambda$ or $\theta_i$ are infinite. This regime is demanding for volumetric time-domain solvers, because the grazing order carries energy parallel to the grating and decays slowly in time. It is also demanding for any method whose discretization is not adapted to the branch point. In the SIM, the anomaly enters only through the spectral part of the Ewald sum, Eq. \eqref{eq:Gspec}, which is evaluated in closed form for every $\kappa_n\neq0$.

\begin{figure}[h]
\centering
\includegraphics[width=0.5\linewidth]{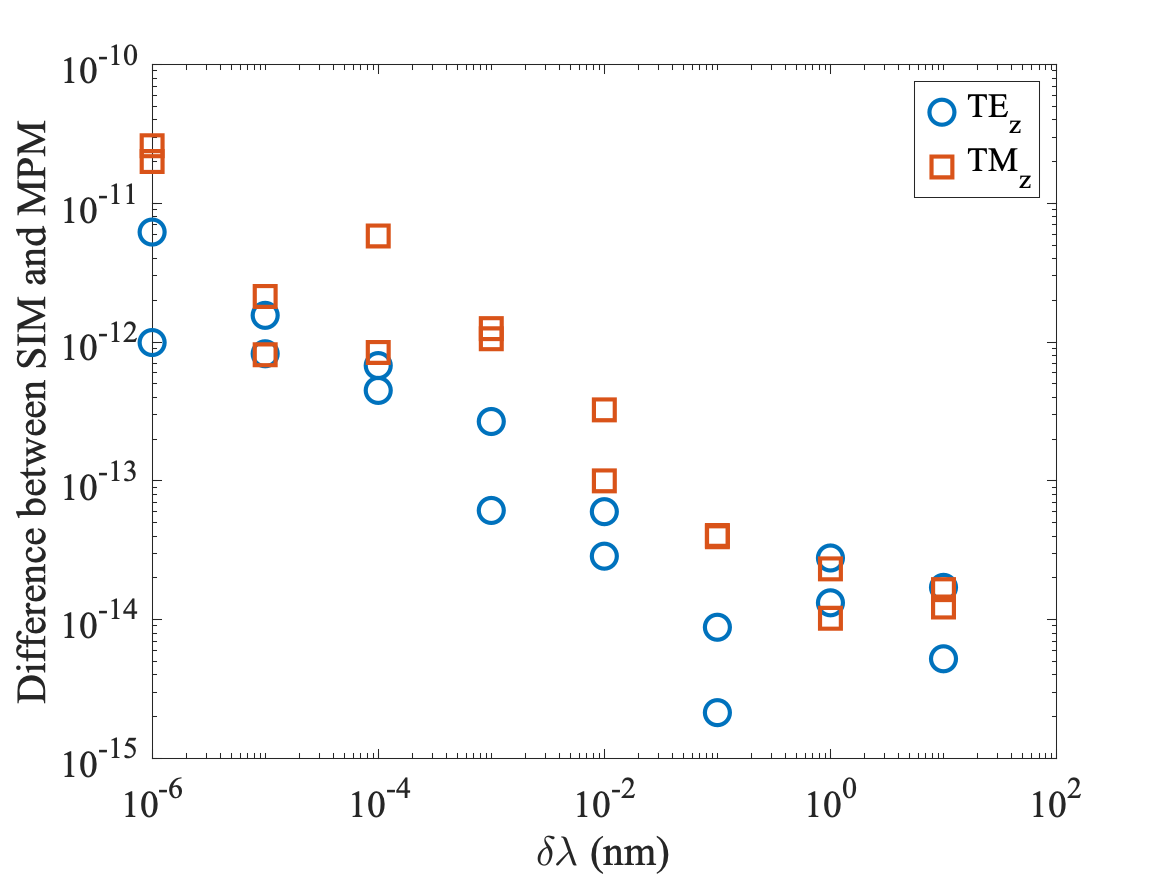}
\caption{Deviation between SIM ($N=256$) and the multipole method versus distance from the $(-1)$ Rayleigh anomaly, $\delta\lambda = \lambda - \lambda_{\rm RA}$. $a=120$ nm, $d=400$ nm, $\theta_i=30^\circ$, $\lambda_{\rm RA}=600$ nm.}
\label{fig:wood}
\end{figure}

The test case is the gold-wire grating with $a=120$ nm, $d=400$ nm, $\theta_i=30^\circ$. For these values the $(-1)$ order is grazing at exactly $\lambda_{\rm RA}=d(1+\sin\theta_i)=600$ nm. Figure~\ref{fig:wood} compares the SIM ($N=256$) with the multipole solution as $\lambda$ approaches $\lambda_{\rm RA}$ from both sides, down to $10^{-6}$ nm, where $|\kappa_{-1}|d=3\times10^{-4}$. The agreement is $\le4\times10^{-13}$ for $|\lambda-\lambda_{\rm RA}|\ge10^{-3}$ nm. It degrades only mildly as $1/|\kappa_{-1}|$ grows, reaching $1.7\times10^{-12}$ (TE$_z$) and $1.8\times10^{-11}$ (TM$_z$) at $10^{-6}$ nm. The results for $N=128$ and $N=256$ coincide at the same level, and the energy residual stays below $10^{-12}$. 

\subsection{Compute Time}
\begin{figure}[h]
\centering
\includegraphics[width=0.5\linewidth]{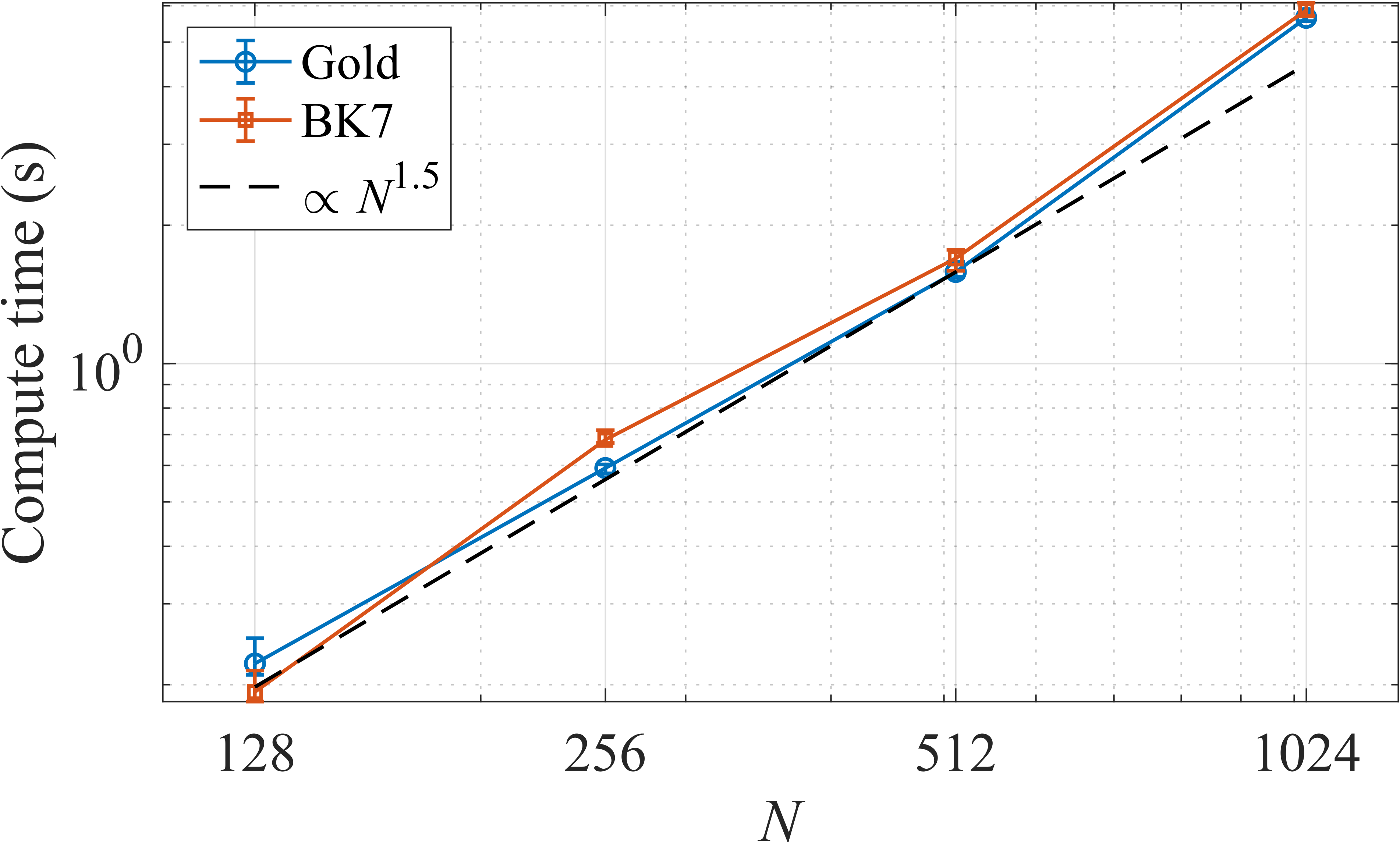}
\caption{Compute time of the SIM versus the number of boundary points $N$ for gratings of gold and BK7 cylinders ($a=\lambda$, gap $\lambda/2$, $\lambda=633$\,nm, TE$_z$, normal incidence). Symbols: median of ten runs; error bars: minimum and maximum. Dashed line: $N^{1.5}$ reference.}
\label{fig:compute_time}
\end{figure}

Figure~\ref{fig:compute_time} shows the compute time of the MATLAB implementation as a function of the number of boundary points $N$ for gratings of gold and BK7 cylinders with radius $a=\lambda$ and gap $\lambda/2$ ($d=2.5\lambda$) at $\lambda=633$\,nm and normal incidence in TE$_z$. Each point is the median of ten runs after one untimed warm-up run, and the error bars show the minimum and maximum. For both materials, the energy residual stays below $5\times10^{-13}$ for all $N$. Over $N=128$--$1024$ the time grows approximately as $N^{1.5}$; the fitted exponents are $1.54$ for gold and $1.61$ for BK7. This is a preasymptotic behavior associated with vectorization, multithreading, and the relative cost of Ewald-kernel evaluation and dense linear algebra; it should not be interpreted as an asymptotic complexity estimate.

\subsection{Beyond the multipole method: lamellar metallic grating}\label{sec:ribbon}
\begin{figure}[h]
\centering\includegraphics[width=0.7\linewidth]{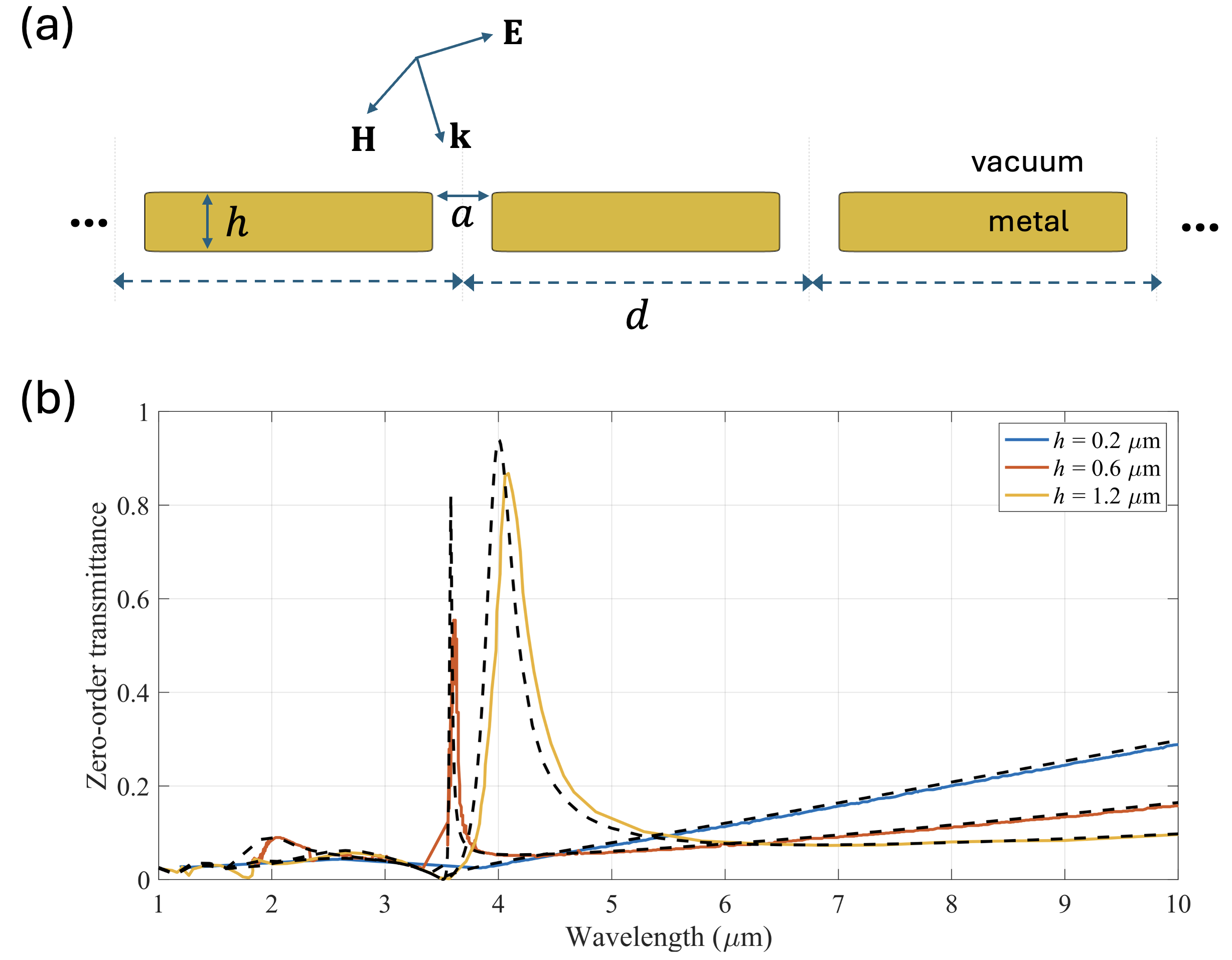}
\caption{(a) Free-standing lamellar metallic grating: bars of height $h$, slits of width $a=0.5\,\mu$m, period $d=3.5\,\mu$m, vacuum, TE$_z$ illumination at normal incidence. (b) Zero-order transmittance for $h=0.2$, $0.6$ and $1.2\,\mu$m. Colored curves: Porto \emph{et al.} \cite{porto}; dashed curves: SIM ($p=32$, $N=1024$, Drude gold \cite{ordal}).}
\label{fig:porto}
\end{figure}

As a final example, we consider a structure whose cross-section is far from circular: the free-standing lamellar metallic grating of \cite{porto}, a standard test case for resonant transmission through subwavelength slits. Metal bars of width $3\,\mu$m and height $h$ are separated by slits of width $a=0.5\,\mu$m, with period $d=3.5\,\mu$m, in vacuum (Fig.~\ref{fig:porto}a). The grating is illuminated at normal incidence with the magnetic field parallel to the slits. We compute the zero-order transmittance $|T_0|^2$ for $h=0.2$, $0.6$ and $1.2\,\mu$m over $1$--$15\,\mu$m. For $\lambda>d$, only the zero order propagates, so $|T_0|^2=\mathcal T$.

The SIM requires a smooth boundary, and the field near a sharp metal corner is singular, so the rectangular bar is replaced by the superellipse $|x/A|^p+|y/B|^p=1$ with $A=(d-a)/2=1.5\,\mu$m and $B=h/2$. It is described in polar form, $\rho(t)=\big[(|\cos t|/A)^p+(|\sin t|/B)^p\big]^{-1/p}$, with an even exponent so that the curve is analytic, and is evaluated in logarithmic form to avoid overflow at large $p$. With $p=32$, the curve coincides with the rectangle except near the corners as shown in Fig. \ref{fig:porto} (a); it encloses $99.85\%$ of the rectangle's area, and its smallest radius of curvature, at the corners, is $7.3$, $20.7$ and $37.1$ nm for $h=0.2$, $0.6$ and $1.2\,\mu$m.

We calculated the transmittance spectra these three gratings for $1~\mu$m$<\lambda<15~\mu$m using  $N=1024$ and $N=2048$. The maximum difference between the two sets of solutions was $6.5\times10^{-5}$ for $h=0.2\,\mu$m, where the node spacing ($6.3$ nm) is comparable to the corner radius. The energy residual is below $4\times10^{-8}$ at $N=1024$ and below $2\times10^{-10}$ at $N=2048$.

Figure~\ref{fig:porto}(b) compares the SIM (dashed black lines) with the results of \cite{porto} (colored curves). The thinnest grating, $h=0.2\,\mu$m, shows no resonance, and its transmittance rises monotonically with wavelength to $0.49$ at $15\,\mu$m. For $h=0.6\,\mu$m a narrow peak appears just above the Rayleigh anomaly at $\lambda=d$, at $3.58\,\mu$m, with a full width at half maximum of about $20$ nm. For $h=1.2\,\mu$m the resonance is broader, about $0.3\,\mu$m wide, and reaches $|T_0|^2=0.94$ near $4\,\mu$m. This evolution is consistent with the two transmission mechanisms identified in \cite{porto}, coupled surface plasmon polaritons on the two faces of the grating and waveguide resonances in the slits. The SIM reproduces the resonance positions, the weak features near $\lambda=d/2$ and the long-wavelength tails of all three curves. The rounded corners and the difference between the Drude model and the metal data of \cite{porto} also shift the resonances slightly, but this should not overshadow the important outcome of this study: the proposed method can be used to study the plasmonic behaviors of metallic gratings.


\section{Conclusions}
The spectral integral method has been extended from a single cylinder to one-dimensional gratings of identical, arbitrarily shaped, smooth cylinders made of lossless dielectrics or lossy and plasmonic metals. It discretizes only one boundary and converges root-exponentially. The logarithmic singularity is removed analytically in the Fourier domain, so that no specialized quadrature is required. The periodicity enters only through the regular part of the quasi-periodic Green's function, which is evaluated with the Ewald method and the Faddeeva function and is therefore valid for any angle of incidence and for complex arguments. Combined with the windowed singularity subtraction, this keeps the treatment of the singular kernels unchanged from the single-cylinder case, even when the interior field of a metal is confined to a thin skin layer. The TE$_z$ and TM$_z$ polarizations share a single formulation, and the Floquet amplitudes of all propagating orders, together with the reflectance, transmittance, absorptance and the residual of the energy balance, follow directly from the boundary data. The SIM remains accurate very close to Rayleigh--Wood anomalies, where the quasi-periodic Green's function diverges. The number of boundary points required is governed mainly by the gap between neighboring cylinders and follows a simple linear rule in the ratio of the perimeter to the gap. Because the cross-section needs only to be smooth and to fit within one period, the method also treats shapes that are far from circular; approximating rectangular bars by high-order superellipses, it reproduces the resonant transmission of a classic lamellar metallic grating.

\section*{Disclosures}
The author declares no conflicts of interest.

\section*{Data availability}
All codes and data required to reproduce the results of this paper are openly available at \url{https://github.com/simsekergun/1D_Periodic_SIM/}.

\bibliographystyle{IEEEtran}
\bibliography{refs_sim1dperiodic}

\end{document}